\documentstyle[12pt]{article}
\date{September 5, 1998}
\title{
Comment on the absence of Coulomb effects on $e^+e^-$ pair
production in ultrarelativistic heavy-ion collisions }
\author{
D.~Yu.~Ivanov$^{1}$\thanks{Email address: d-ivanov@math.nsc.ru} ,
A.~Schiller$^2$\thanks{Email address:
schiller@tph204.physik.uni-leipzig.de} ,
and
V.~G.~Serbo$^{3}$\thanks{Email address: serbo@math.nsc.ru}
\\
{\it $^1$Institute of Mathematics, Novosibirsk, 630090, Russia}\\
{\it $^2$Institut f\"ur Theoretische Physik and NTZ,
Universit\"at Leipzig,}\\{\it D-04109 Leipzig, Germany} \\
{\it $^3$Novosibirsk State University, Novosibirsk, 630090,
Russia}
}

\begin{document}


\maketitle


\begin{abstract}
In recent works  \cite{SW}-\cite{ERSG} it was claimed that the
Coulomb correction to the $e^+e^-$ production cross section in
relativistic heavy-ion collisions is absent. We point out that
this statement is in obvious contradiction to some well known
results. We obtain large Coulomb corrections in the pair
production cross section for the RHIC and LHC heavy-ion
colliders.
\end{abstract}


Two new large colliders with relativistic heavy nuclei, RHIC and
LHC, are scheduled to be in operation in the nearest future. The
charge number of the nuclei $Z_1=Z_2=Z$ and their Lorentz factors
$\gamma_1=\gamma_2=\gamma$ are
$Z=79$ and $\gamma =108$ for RHIC (Au-Au collisions) and
$Z=82$ and $\gamma =3000$ for LHC (Pb-Pb collisions).
One of the important processes at these colliders is
\begin{equation}
 Z_1Z _2\to Z_1Z_2 e^+e^- \,.
 \label{0}
 \end{equation}
Its cross section is huge and in the Born approximation
the total cross section is equal to \cite{LL,R}
\begin{equation}
\sigma_{\mathrm{Born}} = {28\over 27\pi}\,
{Z^4\alpha^4 \over m_e^2}\,
\left( L^3 - {178\over 28}
L^2 + {7\pi^2 + 370\over 28}L - 13.8 \right)\,,\;\;
L= \ln{(4\gamma^2)}\,.
\label{1}
\end{equation}
Therefore,  $\sigma_{\mathrm{Born}} = 36 $ kb for RHIC and 227
kb for LHC and this process should be considered as  a serious
background for a number of planned experiments.
Besides, the process (\ref{0})
is also important for the determination of the beam lifetime.
The pair production, accompanied by the capture of an electron on
the atomic orbit, is the leading beam loss mechanism. It means
that the differential and total cross sections of this process
should be known with a good accuracy.


Since in heavy-ion collisions the expansion parameter $Z\alpha
\sim 1$, the whole series in this parameter has to be summed in
order to obtain the correct cross section.  Following
Ref.~\cite{BM}, we define the Coulomb correction (CC) as the
difference $d\sigma_{\mathrm{Coul}}$ between the whole sum
$d \sigma$ and
the Born approximation
\begin{equation}
d\sigma = d \sigma_{\mathrm{Born}} + d \sigma_{\mathrm{Coul}}\,.
\label{2}
\end{equation}


Recently, in Refs.~\cite{SW}-\cite{ERSG} these Coulomb effects
were studied using an approximate solution of the Dirac equation.
These authors argue that the complete result (\ref{2}) coincides
with that of the Born contribution called perturbative result. In
other words, the whole amplitude of process (\ref{0}) is equal to
the Born amplitude up to a phase factor depending on $Z\alpha$.
In those papers the symmetric case $Z_1=Z_2$ was considered, but
the authors of \cite{BL,ERSG} also argue that their results are
trivially extended to the asymmetric case $Z_1\neq Z_2$. In the
latter case their conclusion about this amplitude behavior
remains valid.


In this comment we point out that these conclusions are in
contradiction with well known results about CC.


Remind that in the photoproduction of $e^+e^-$
pairs on nuclei (see Refs.~\cite{BM}, \S 32.2 of~\cite{AB} and
\S 98 of~\cite{BLP}) the Coulomb correction to the total cross
section decreases the Born contribution by about 10 \% for a Pb
target.  It is worthwhile to note that the correct result is
obtained only taking into account $1/\gamma$ corrections in the
asymptotic expressions for the lepton wave functions. In a recent
paper \cite{IM} these results have been reproduced summing the
whole perturbative series in $Z \alpha$.


For the lepton pair production in the collisions of two nuclei
with $Z_1\alpha \ll 1$ and $ Z_2 \alpha \sim 1$ the Coulomb
correction  has been obtained in Refs.~\cite{NP,BB}. Recently, we
and Kuraev \cite{IKSS} have calculated  CC for the $e^+e^-$ pair
production in the collisions of muons with heavy nuclei.  All
mentioned results agree with each other and  noticeably change
the Born cross sections.


The most transparent example is the collision
of ultrarelativistic light ($Z_1\alpha \ll 1$) and heavy ions
in the region of small momentum transfer squared for the light
ion $-q_1^2\ll m_e^2$.  In this case the cross section (\ref{2})
is given as the product of the number of equivalent photons
generated by the light ion $d n_{\gamma/Z_1}(\omega, q_1^2)$ with
the cross section for the process $\gamma Z_2 \to Z_2 e^+e^- $:
\begin{equation}
d \sigma_{Z_1Z_2\to Z_1Z_2 e^+e^-} = d
n_{\gamma/Z_1}(\omega,q_1^2) d \sigma_{\gamma Z_2 \to
Z_2e^+e^-}(\omega)\,,
\label{3}
\end{equation}
$\omega$ is the energy of the equivalent photons.
We want to stress that $d \sigma_{\gamma Z_2 \to
Z_2e^+e^-}(\omega)$ calculated in Ref.~\cite{BM} differs
considerably from the Born cross section for photoproduction and,
therefore, the cross section (\ref{2}) differs from $d \sigma_{\mathrm{Born}}$.


Based on the well grounded approach used in Ref.~\cite{IKSS} we
have calculated the Coulomb correction to the $e^+e^-$ pair
production in relativistic heavy-ion collisions \cite{ISS}.  With
an accuracy of the order of $1 \%$ we find
\begin{equation}
\sigma_{\mathrm{Coul}}/ \sigma_{\mathrm{Born}} = 25\, \% \;\;
{\mathrm for \ \ RHIC}\,, \ \ \
\sigma_{\mathrm{Coul}}/ \sigma_{\mathrm{Born}} = 14\, \% \;\;
{\mathrm for \ \ LHC}
\label{4}
\end{equation}
in clear contradiction to \cite{SW}-\cite{ERSG}.


\section*{Acknowledgments}
We are very grateful to G.~Baur, V.~Fadin, I.~Ginzburg and
Yu.~Dokshitzer for useful discussions.



\end{document}